\documentclass[final,5p,times,twocolumns,10pt]{elsarticle}

\usepackage[english]{babel}

\usepackage{graphics}
\usepackage{amssymb,amsthm,amsmath}
\usepackage{booktabs}
\usepackage{multirow}

\usepackage[T1]{fontenc}
\usepackage[utf8]{inputenc}
\usepackage{lmodern}

\graphicspath{{./figs/}}

\journal{NIM A  RICAP-2013}

\begin{document}
\begin{frontmatter}

\title{Radio constraints on Galactic WIMP dark matter}

\author{Roberto A. Lineros}

\address{Instituto de F\'{\i}sica Corpuscular (CSIC-Universitat de Val\`{e}ncia), Apdo. 22085, E-46071 Valencia, Spain.}

\ead{rlineros@ific.uv.es}

\begin{abstract}
Synchrotron emission from electron cosmic ray populations can be used to study both cosmic rays physics and WIMP dark matter imprints on radio skymaps. We used available radio data - from MHz to GHz - to analyze the contribution from galactic WIMP annihilations and impose constraints on WIMP observables: annihilation cross section, channel and mass. Depending on the annihilation channel we obtain as competitive bounds as those obtained in FERMI-LAT gamma ray analysis of dwarf satellite galaxies.
\end{abstract}


\end{frontmatter}


\section{Introduction}
\label{sec:intro}

During the last decades improvements on cosmological observations have confirmed that around 27\% of the matter content of observable Universe is dominated by non-barionic Dark Matter (DM)~\cite{Planck:2013}. 
The most popular DM candidates are in the form of Weakly Interactive Massive Particles (WIMPs) due to their diverse and complementary ways of detection.
Moreover, WIMPs are common to many particle physics models beyond the Standard Model (SM) e.g. Supersymmetry, extra-dimensions, and extended Higgs sector models.
All these candidates share similar characteristics: non-colored, electrically neutral and stable on cosmological scales.
WIMPs participate actively in the early universe until they freeze out from the thermal bath due to the expansion of the universe.
This mechanism provide a natural way to reproduce the observed relic abundance~\cite{Planck:2013}, 
\begin{equation}
\Omega_{\rm DM} h^2 = 0.1196 \pm 0.0031 \, , 
\end{equation} 
and it requires cross section of the order of electroweak interactions.
This cross section is usually referred as the thermal WIMP cross section: 
\begin{equation*}
\langle \sigma v \rangle = 3 \times 10^{-26} \, {\rm cm}^3 {\rm s}^{-1}. 
\end{equation*}

One of the expected signatures are the annihilation products from WIMPs around celestial objects like the Milky Way and its dwarf satellite galaxies.
The SM particles coming from WIMP annihilations can produce gamma rays, cosmic rays, and neutrinos.
Current and projected observatories can explore these new signals thanks to their high sensitivity and resolution. 
For example, this is the case of the FERMI-LAT gamma ray observatory.
On the other hand, radio surveys have been continuously observing and characterizing the galactic activity. 
One of the most known all--sky survey is the HASLAM~map~\cite{1982A&AS...47....1H} at the frequency of 408~MHz and it is widely used in astrophysics.\\

This work is based on Ref.~\cite{2012JCAP...01..005F} and describes its main results.

\section{Constraints from radio data}

We consider the synchrotron emission at radio frequencies, i.e. from MHz to GHz, from electrons and positrons as result of WIMP annihilations of the DM halo of the Milky Way. 
In order to cover the whole possibilities, we take into account the uncertainties regarding the production of synchrotron emission:
\begin{itemize}
	\item[(i)] Galactic magnetic field (GMF) ranging between 1--10 $\mu$G.
	\item[(ii)] Physically motivated GMF spatial distributions.
	\item[(iii)] Cosmic ray propagation models.
	\item[(iv)] DM distributions: isothermal and NFW profile~\cite{1996ApJ...462..563N}.
	\item[(v)] DM annihilation into: muons, taus, $W$~bosons and $b$~quarks.
	\item[(vi)] DM annihilation cross section similar to the thermal value.
\end{itemize}
For most of these scenarios, the radio emission coming from WIMPs is as bright as the available radio surveys.
Thus, using the radio survey we set constraints for some of the DM properties e.g. the annihilation cross section.\\

The bounds are then obtained by requiring that the DM signal would not exceed the observed radio emission by 3 standard deviations, i.e.
\begin{equation}
	T_{\rm DM}(\nu) \le T_{\rm obs}(\nu) + 3 \,\sigma_{\rm survey}(\nu) \, .
\end{equation}
The radio surveys and the DM emision skymap where previously divided into patches of $\sim 10^{\circ} \times 10^{\circ}$ and then compared.
We did not include any astrophysical radio sources meaning that our bounds are conservative.\\

\begin{figure}[tb]
\centering
\includegraphics[width=0.99\columnwidth]{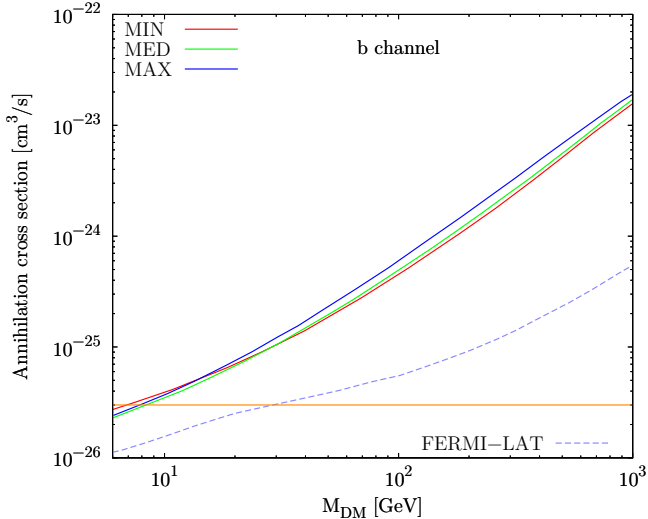}
\caption{Upper bounds on the annihilation cross section versus DM mass for annihilations into $b$. In dashed line, we show bounds obtained by FERMI-LAT from the analysis on dwarf satellite galaxies~\cite{2011PhRvL.107x1302A}. Radio constraints are less competitive with respect to hadronic channels. Further details in Ref.~\cite{2012JCAP...01..005F}.
\label{fig:1}}
\end{figure}

\begin{figure}[tb]
\centering
\includegraphics[width=0.99\columnwidth]{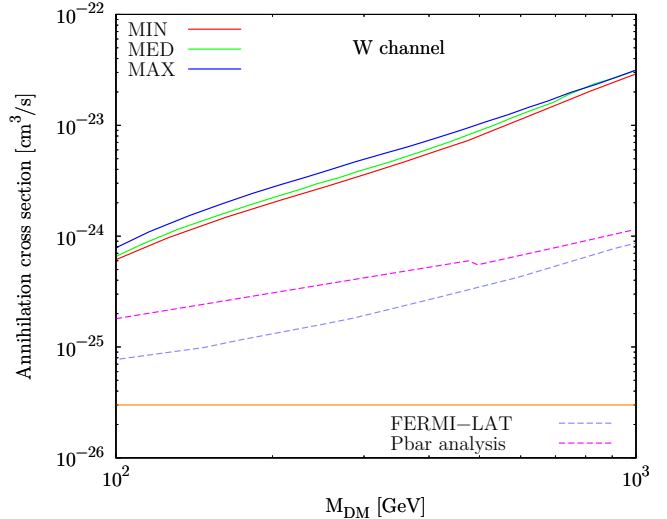}
\caption{Upper bounds on the annihilation cross section versus DM mass for annihilations into $W$. Blueish dashed line are the bounds from FERMI-LAT analysis, also we show the bounds from antiproton/proton analysis~\cite{2009PhRvL.102g1301D}. Further details in Ref.~\cite{2012JCAP...01..005F}.
\label{fig:2}}
\end{figure}

\begin{figure}[tb]
\centering
\includegraphics[width=0.99\columnwidth]{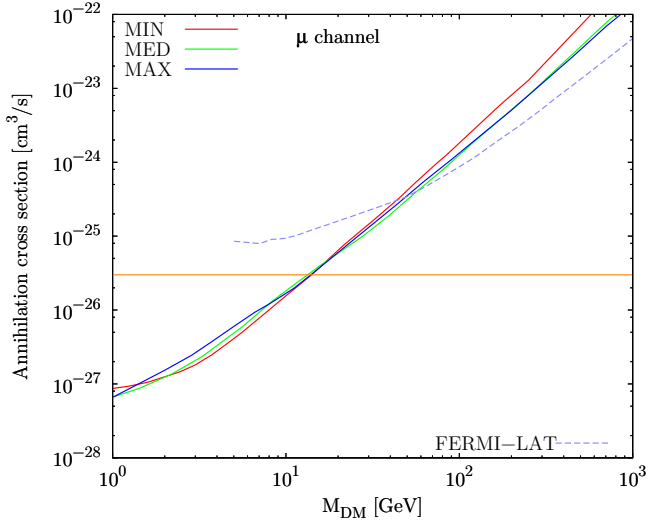}
\caption{Upper bounds on the annihilation cross section versus DM mass for annihilations into muons. Radio constraints extracted from leptonic channel are in better situation than constraints obtained by FERMI-LAT. Further details in Ref.~\cite{2012JCAP...01..005F}.
\label{fig:3}}
\end{figure}

This procedure mainly constrains the annihilation cross section for a specific annihilation channel.
In Figs.~\ref{fig:1}, \ref{fig:2} and \ref{fig:3},  we present some representative results of the analysis carried in Ref.~\cite{2012JCAP...01..005F}.
All of these show upper bounds on the annihilation cross section for different annihilation channels versus the WIMP mass. 
In Fig.~\ref{fig:1}, we analyze the case of annihilation into $b$~quarks. We note that our bounds are less restrictive than those obtained in the gamma ray analysis of dwarf satellite galaxies~\cite{2011PhRvL.107x1302A}. This is mainly due to annihilations into quarks produce much more gamma rays because of large $\pi^0$ production and subsequent decay. 
Similar situation happens in Fig.~\ref{fig:2} when we compare our results with previous works on constraints from antiproton/proton observation~\cite{2009PhRvL.102g1301D}.\\

The opposite occurs when we compare the muon channel, Fig.~\ref{fig:3}. In this case, radio bounds are more competitive than those from ~\cite{2011PhRvL.107x1302A} because muonic annihilation channel (and in general any leptonic channel) produce larger amount of electrons and positrons.\\

For the $b$ and muon channels, the bounds reach the thermal cross section value at low WIMP masses. This method constrains particle physics models with predicted annihilation channels and cross section higher than our bounds.\\

Our analysis shows the complementarity between radio and gamma rays observations for the study of galactic DM. In principle, a combined analysis in this direction would improve DM indirect searches.\\

\section{Conclusions}

Synchrotron emission from galactic WIMP annihilations presents an interesting and alternative observable to study the WIMP properties.
This method is specially efficient for constraining annihilation into leptons. Our bounds are quite competitive with respect to similar analysis and reach the value of WIMP thermal cross section for WIMP masses lower than 15 GeV.

\section*{Acknowledgements}

We acknowledge N. Fornengo, M. Regis and M. Taoso.
This work was supported by the Spanish MINECO under grants \mbox{FPA2011-22975} and MULTIDARK \mbox{CSD2009-00064} (Consolider-Ingenio 2010 Programme). Also, it was supported by Prometeo/2009/091 (Generalitat Valenciana), and by the EU ITN UNILHC \mbox{PITN-GA-2009-237920}.


\newcommand{\apj}{Astrophys. J.}
\newcommand{\apjl}{Astrophys. J. Lett.}
\newcommand{\apjs}{Astrophys. J. Suppl. Ser.}
\newcommand{\aap}{Astron. \& Astrophys.}
\newcommand{\aj}{Astron. J.}
\newcommand{\araa}{Ann. Rev. Astron. Astrophys. } 
\newcommand{\mnras}{Mon. Not. R. Astron. Soc.}
\newcommand{\physrep}{Phys. Rept.}
\newcommand{\jcap}{JCAP}
\newcommand{\prl}{PRL}
\newcommand{\prd}{Phys. Rev.}

\bibliographystyle{elsarticle-num}
\bibliography{refs.bib}

\end{document}